\documentclass[a4paper,fleqn,usenatbib]{mnras}
\usepackage{graphicx}
\usepackage{amssymb}
\usepackage{amsmath}
\usepackage{times} 
\usepackage{multirow}
\usepackage{url}
\hypersetup{draft} %****** remove

\title[The angular scale of homogeneity in the Local Universe]{The angular scale of homogeneity 
in the Local Universe with the SDSS blue galaxies}

\author[Avila et al.]{F. Avila,$ ^{1} $\thanks{e-mail: felipeavila@on.br}
C. P. Novaes,$^{1}$
A. Bernui,$^{1}$
E. de Carvalho,$^{1,2}$
J. P. Nogueira-Cavalcante$ ^{1} $\\
$^{1}$Observat\'orio Nacional, Rua General Jos\'e Cristino 77, 
S\~ao Crist\'ov\~ao, 20921-400 Rio de Janeiro, RJ, Brazil \\
$^{2}$Centro de Estudos Superiores de Tabatinga, Universidade do Estado do Amazonas, 
69640-000, Tabatinga, AM, Brazil \\
}

\date{Accepted XXX. Received YYY; in original form ZZZ}

\pubyear{2015}

\begin{document}

\maketitle

\begin{abstract}
We probe the angular scale of homogeneity in the local Universe using blue galaxies from 
the SDSS survey as a cosmological tracer. 
Through the scaled counts in spherical caps, $ \mathcal{N}(<\theta) $, and the fractal correlation 
dimension, $\mathcal{D}_{2}(\theta)$, we find an angular scale of transition to homogeneity for this 
sample of $\theta_{\text{H}} = 22.19^{\circ} \pm 1.02^{\circ}$. 
A comparison of this measurement with another obtained using a different cosmic tracer at a similar 
redshift range ($z < 0.06$), namely, the HI extragalactic sources from the ALFALFA catalogue, 
confirms that both results are in excellent agreement (taking into account the corresponding bias 
correction). 
We also perform tests to asses the robustness of our results. 
For instance, we test if the size of the surveyed area is large enough to identify the transition scale 
we search for, and also we investigate a reduced sample of blue galaxies, obtaining in both cases 
a similar angular scale for the transition to homogeneity. 
Our results, besides confirming the existence of an angular scale of transition to homogeneity in 
different cosmic tracers present in the local Universe, show that the observed angular scale 
$\theta_{\text{H}}$ agrees well with what is expected in the $\Lambda$CDM scenario. 
%according to simulations. 
Although we can not prove spatial homogeneity within the approach followed, our results provide 
one more evidence of it, strengthening the validity of the Cosmological Principle. 
\end{abstract}
%Our work strengthens the validity of the cosmological principle therefore, the $\Lambda$CDM 
%model, for the local Universe in the model independent way.


%%-------------------------------------------------------------------------------------------------------------------------------
\section{Introduction}\label{sec1}

In few years cosmology has achieved an unprecedented level in the restriction of parameters of 
the most reliable cosmological models due, mainly, to the high quality of observational data and 
to the diversity of cosmic tracers observed~\citep{Planck2018,DES,Alam2017}. 
Nowadays, these data is being used to test the Cosmological Principle (CP), the main hypothesis 
that supports the concordance model of cosmology.

The statistical isotropy of the universe is being tested for many extragalactic objects like 
Radio sources~\citep{Blake2002,Ghosh2016}, 
Gamma-ray bursts~\citep{BFW,Tarno2017,Ripa2018}, 
galaxy clusters~\citep{Bengaly2017a}, and galaxy datasets like the WISE~\citep{Yoon,%
Bengaly2017b,Camila2018} and the SDSS catalogues~\citep{Sarkar19}, 
where all these analyses show a good concordance with the isotropy of the universe. 
The Planck Convergence and Cosmic Microwave Background temperature fluctuations maps 
have been examined and are also consistent with statistical isotropy at small angular 
scales~\citep{Ade2016,Camila2016,Marques2018}, although some controversy remains at large 
angles~\citep{Bernui2008,Gruppuso2013,Polastri2015,%
Schwarz2016,Aluri2017,Rath2017,Bernui2018}.

The study of spatial homogeneity using datasets is more recent, and is a delicate issue. 
Methods that explore the homogeneity of the matter distribution counting objects in spheres or 
spherical caps (in case the objects are projected on the sky) are not direct tests of spatial 
homogeneity because the counts are restricted to the intersection of the past light-cone with the 
spatial hyper-surfaces \cite[see, e.g.,][]{Ellis06,Maartens11}. 
Rigorously, these methods provide consistency tests in the sense that: if the counting methods 
shows that the objects distribution does not approach 
homogeneity on any scale, then this can falsify the CP. 
On the other hand, if data confirm the existence of a transition scale to homogeneity, then this 
strengthens the evidence for spatial homogeneity, but does not prove it~\citep{Maartens11}.

To investigate homogeneity in a given sample, one searches for a characteristic scale, 
$r_{\text{H}}$, beyond which the distribution of cosmic objects can be considered homogeneous. 
In the \textit{counts-in-spheres} method, one averages the number of objects inside spheres 
of radius $r$, $N(< r)$, and take their logarithmic derivative to obtain the fractal correlation 
dimension, $D_{2}(r)$~\citep{Scrimgeour}. 
For a homogeneous distribution of objects, one should expect a behaviour like 
$N(< r) \propto r^{D_{2}}$, with $D_{2} < 3$ at small scales, $r < r_{\rm H}$, and $D_{2} = 3$ at 
larger sphere radius, $r > r_{\rm H}$~\citep{Castagnoli,Sarkar2018}. 
However, this method does not deal correctly with incomplete catalogues and boundary effects; 
remember that often the astronomical data is located in disconnected sky patches, with contours 
that are not straight lines, and may contain holes due to the application of masks. 
To solve these inconveniences,~\cite{Scrimgeour} introduced the {\em scaled counts-in-spheres} method, which consists of taking a normalised ratio of counts, considering the data and simulated homogeneous samples. 
This method and the equivalent quantity adapted to a 2-dimensional (2D) analysis, the \textit{scaled counts-in-caps} (or \textit{spherical caps}), have been applied to study homogeneity in several catalogues 
[\cite{Nadathur,Laurent,Pandey2015,Sarkar,Ntelis,Rodrigo,Rodrigo18}; for a different approach 
see, e.g., \cite{Hoyle}]. 

The present work analyses the clustering features of the current sample of blue galaxies from the 
Sloan Digital Sky Survey (SDSS). 
The three main objectives of doing so, are: 

\noindent
(i) to study if an angular scale of transition to homogeneity can be revealed by the SDSS blue 
galaxies in the local Universe; 

\noindent
(ii) to find the relative bias between two cosmic tracers (at similar redshift): the HI sources from 
the ALFALFA catalogue and the SDSS blue galaxies, 
$b_{\,\mbox{\footnotesize blue\,/\,\sc{HI}}}$; 

\noindent
(iii) to confirm if the scale found in (i) equals --considering the relative bias-- the angular scale of 
homogeneity in the local Universe recently found by~\cite{Felipe} using the ALFALFA catalogue 
\cite[homogeneity scales obtained in two samples can be compared if one knows the relative bias; see, e.g.,][]{Scrimgeour,Ntelis}. 

Our work is organised as follows. In section 2 we present the selection of the data. In section 3 we describe the \textit{scaled counts-in-caps} (SCC) method and how it is used to estimate the angular scale of homogeneity. 
In sections 4 and 5 we show our results and present our conclusions, respectively.

%%----------------------------------------------------------------------------------------------------------------------------
%Original: 
%35356 blue galaxies, in the same redshift interval $z \in [0,0.06]$ as 
%the ALFALFA HI sources, the median value here is slightly larger, $\bar{z} = 0.037$ 
%mean    = 0.036991919108496432
%median = 0.037159999999999999
%
%Gaussianizado: 
%21226 blue galaxies with: 0.0 < z < 0.06
%mean    = 0.028982754310750966
%median = 0.028275000000000002
%%---------------------------------------------------------------------------------------------------------------------------
\section{Data}\label{sec2} 

We select blue star-forming galaxies from the galaxy colour-colour diagram, using the $u$, $g$, 
and $r$ Sloan Digital Sky Survey (SDSS) broad bands~\citep{York}. 
The data used is part of the twelfth public data release, DR12, of the SDSS 
collaboration~\citep{Alam2015}. 
The SDSS magnitudes for each galaxy is corrected by Galactic extinction 
following~\citet{Schlegel1998}. 
We apply k-correction using the PYTHON version of the K-correction calculator\footnote{\url{http://kcor.sai.msu.ru/getthecode/}} \citep{Chilingarian2010, Chilingarian2012}.

\begin{figure}
\includegraphics[width=\columnwidth]{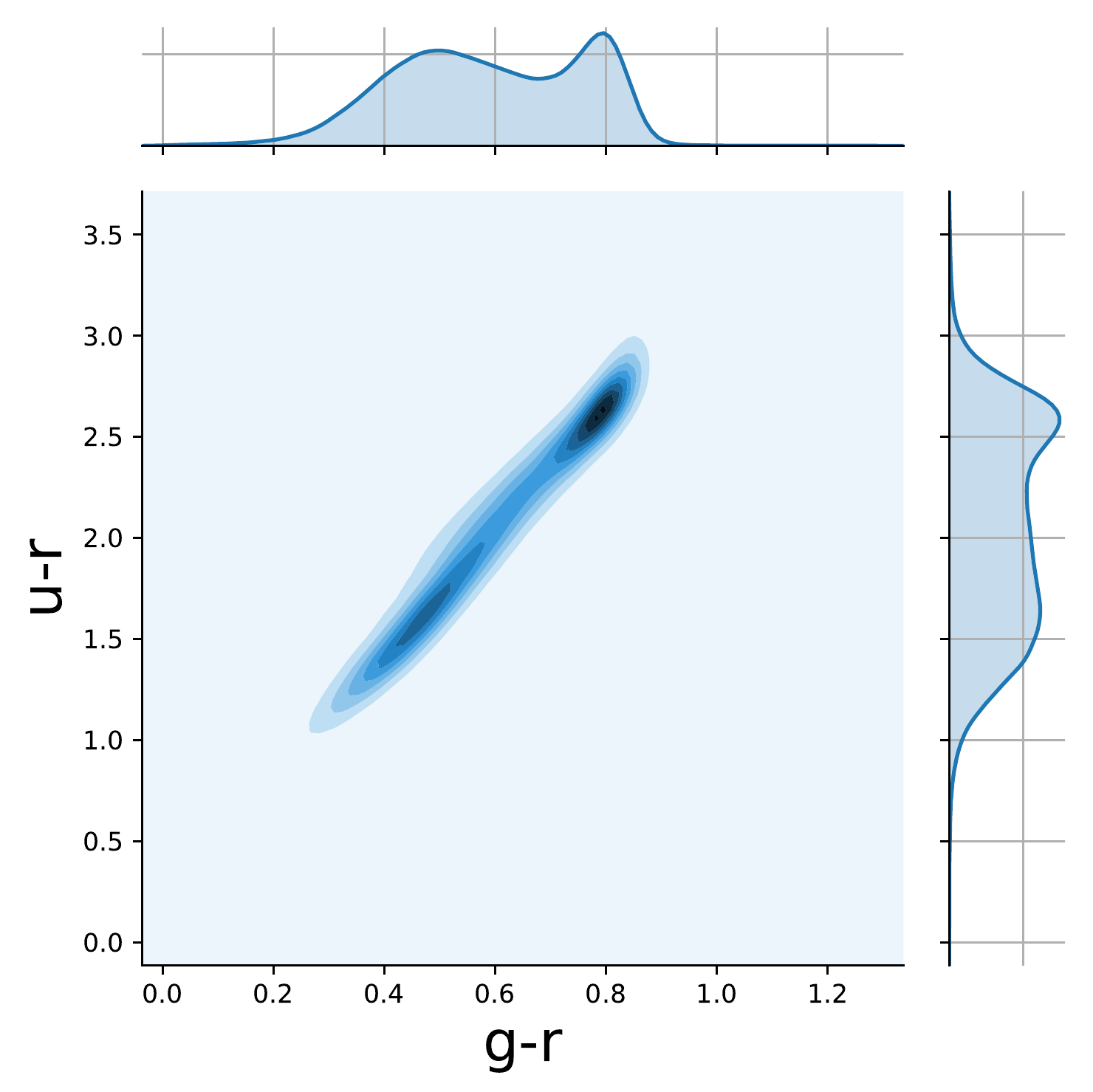}
%\vspace{-0.5cm}
\caption{Colour-colour diagram of the SDSS galaxies, corrected by Galactic and intrinsic extinctions. 
The horizontal and vertical plots show the bimodal distribution of $g-r$ and $u-r$ galaxy colours, 
respectively.}
\label{fig1}
%colour_magnitude_diagram
\end{figure}

Most  blue star-forming  galaxies  are  obscured  by dust, occupying the redder parts of the 
galaxy colour-magnitude diagrams (CMD), both in the local Universe~\citep{Sodre2013} and at 
higher redshifts~\citep{Goncalves2012}. 
We correct $u$, $g$, and $r$ SDSS magnitudes by intrinsic reddening, through the flux of 
H$_{\alpha}$ and H$_{\beta}$ emission lines~\citep[measurement details are 
described in ][]{Brinchmann2004}. 
Following~\cite{Calzetti1994}, we determine the intrinsic $B-V$ excess by 
\begin{equation}
E(B-V) = 0.935 \ln\left(\frac{\text{H}_{\alpha}/\text{H}_{\beta}}{2.88}\right) \times 0.44 \, .
\end{equation}
We convert $E(B-V)$ into extinction in SDSS bands following \citet{Calzetti2000} 
\begin{equation}
A_{\lambda} = k_{\lambda} \times E(B-V) \,\, ,
\end{equation}
where
\begin{equation}
k(\lambda)=1.17\left(-2.156+\frac{1.509}{\lambda}-\frac{0.198}{\lambda^2} + \frac{0.011}{\lambda^3}\right) + 1.78 \, ,
\end{equation}
for \,$0.12 \leq \lambda[\mu\text{m}] \leq 0.63$, and 
\begin{equation}
k(\lambda) = 1.17\left(-1.857+\frac{1.040}{\lambda}\right) + 1.78 \, ,
\end{equation}
for \,$0.63 \leq \lambda[\mu\text{m}] \leq 2.20$. 
Figure~\ref{fig1} shows the galaxy colour-colour diagram of our sample. 
We define blue star-forming galaxies those from the region between $0.0 < g-r < 0.6$ and 
$0.0 < u-r < 2.0$ on the diagram.

%%----------------------------------------------------------------------------------------------------------------

\begin{figure}
\centering
\includegraphics[width=\columnwidth]{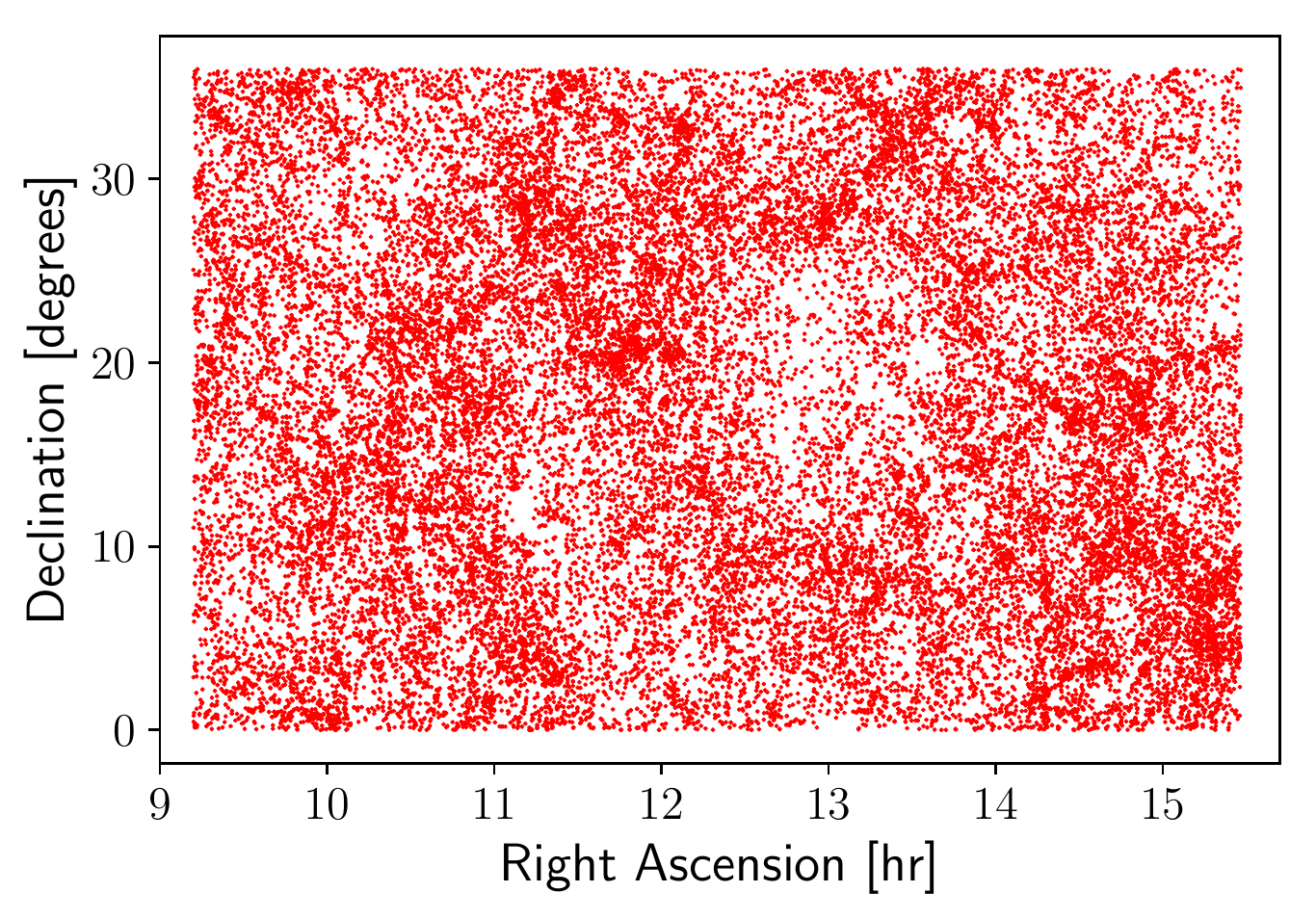}
%\vspace{-0.5cm}
\caption{Cartesian projection of the SDSS sample selected for our analyses; it contains a total of 35356 blue galaxies.}
\label{fig2}
\end{figure}

\begin{figure}
\centering
\includegraphics[width=\columnwidth]{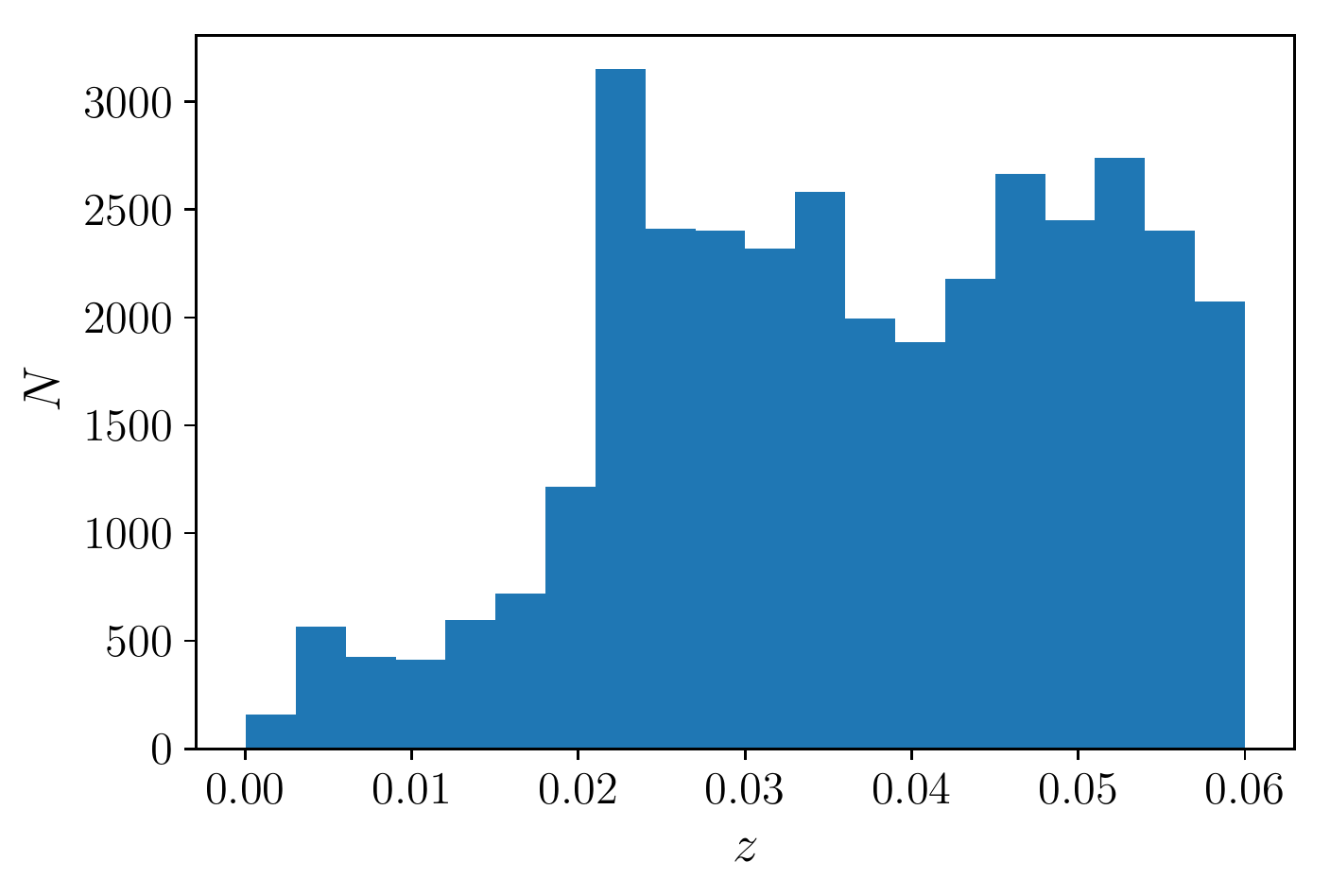}
%\vspace{-0.5cm}
\caption{Redshift distribution of the SDSS blue galaxies sample selected for our homogeneity 
analyses.}
\label{fig3}
\end{figure}

From this sample of SDSS blue galaxies we select those with the same observational features 
used in  \citet[][from now on we call this reference A18]{Felipe}, that is, data in the same sky 
patch and redshift interval (i.e., $z \in [0,0.06]$). 
This means that the selected SDSS blue galaxies have angular coordinates: 
$9^{\text{h}}20^{\text{m}} \leqslant \text{RA} \leqslant 15^{\text{h}}50^{\text{m}}$ and 
$0^{\circ} \leqslant \text{DEC} \leqslant 36^{\circ}$ for Right Ascension and Declination, 
respectively. 
Under these conditions the selected sample has 35356 SDSS blue galaxies with the angular 
distribution shown in figure~\ref{fig2}, while in figure~\ref{fig3} we observe their redshift distribution 
$0.0 < z < 0.06$, with median redshift of $0.037$.

%%-------------------------------------------------------------------------------------------------------------------------
\section{Methodology}\label{sec3} 

In this work we apply the SCC method to search for the angular scale of transitions to homogeneity 
in the projected distribution of the SDSS blue galaxies in the local Universe ($z < 0.06$), using the 
same region of the universe (projected area and redshift range) analysed in A18.
The reason is that one of our main objectives here is to perform comparative analyses to confirm 
the angular scale of homogeneity recently found in A18 using the ALFALFA catalogue (HI extragalactic 
sources). 

In this section we briefly present the SCC method, calculated using the Landy-Szalay estimator, 
and the definition of the fractal correlation dimension, $\mathcal{D}_{2}$ (more details can be found 
in the section 3 of A18). 
We also describe the criterium used to determine the angular scale of transition to homogeneity, 
$\theta_{\rm H}$, for a 2D distribution on a sphere.

%%%---------------------------------------------------------------------------------------------------------------------
\subsection{The scaled counts-in-caps: the Landy-Szalay estimator} \label{sec3p1}

We are interested in the analysis of the 3D data projected on the sky, that is, 
the study of data projected in a 2D region on the celestial sphere; for this we replace the 3D 
spheres of radius $r$ for 2D spherical caps of angular radius $\theta$. 
Then, we can define the scaled counts-in-caps 

\begin{equation}\label{eq1}
\mathcal{N}(<\theta) \,\equiv\, \dfrac{N_{\rm gal}(<\theta)}{N_{\rm rand}(<\theta)} \, ,
\end{equation}
and the fractal correlation dimension 

\begin{equation}\label{eq2}
\mathcal{D}_{2}(\theta) \,\equiv\, \dfrac{d~{\rm ~ln}\mathcal{N}(<\theta)}{d~{\rm ~ln}~\theta} 
\,+\, \dfrac{\theta~\sin\theta}{1 - \cos \theta} \,\, ,
\end{equation}
where $N_{\rm gal}$ is the average counting of sources inside a spherical cap of radius $\theta$. 
The quantity $N_{\rm rand}$ has the same meaning but for a random catalogue, constructed with 
the same footprint and number of objects from the selected sample. 
To measure $ \mathcal{N}(<\theta) $ we chose to use the following estimator 

\begin{equation}\label{eq3}
\mathcal{N}_{j}(<\theta) \,=\, 1 
\,+\, \dfrac{1}{1 - \cos \theta} \int\limits_{0}^{\theta} \omega_{j}(\theta')\sin\theta'd\theta'  \, , 
\end{equation}
where $\omega_j(\theta$) is the Landy-Szalay two-point angular correlation 
function~\citep{LS,Edilson}, calculated for the $j$th random catalogue as\footnote{Analyses using 
other estimators can be found in, e.g.,~\cite{Ntelis,Rodrigo}.} 

\begin{equation}\label{eq4}
\omega_j(\theta) \,=\, \dfrac{DD(\theta) - 2DR(\theta) + RR(\theta)}{RR(\theta)} \, .
\end{equation}
$D\!D(\theta)$ is defined as the number of pairs of galaxies in the data sample, with an angular separation $\theta$, normalised to the total number of pairs, $RR(\theta)$ is a similar quantity but calculated for pairs in the random catalogue, and $D\!R(\theta)$ corresponds to the number of pairs, with one object in the dataset and other in the random catalogue, normalised to the total number of pairs, using as centre the position of the object in the dataset. 

For the $j$th random catalogue we calculate $\omega_{j}(\theta)$ in the range 
$1^{\circ} \leq \theta \leq 40^{\circ}$ in bins of width $\Delta \theta = 1^{\circ}$, perform a best-fit of 
these points to obtain a function to calculate the integral in equation~(\ref{eq3}), and then use 
equation~(\ref{eq2}) to calculate $D_{2}^{j}(\theta)$, with $j = 1,\cdots,20$.

%%%-----------------------------------------------------------------------------------------------------------------
\subsection{The definition of the homogeneity scale}

To investigate homogeneity in a projected sample, one searches for a characteristic scale, 
$\theta_{\text{H}}$, beyond which the distribution of cosmic objects can be considered 
homogeneous. 
The value of this scale, $\theta_{\text{H}}$, is here obtained adopting the 1\%-criterium, to be 
defined below. 
In principle, it seems an arbitrary criterium, but in reality~\cite{Scrimgeour} show that it has been 
obtained through rigorous arguments, and since then it has been adopted in several 
analyses~\citep{Laurent,Ntelis,Rodrigo}. 
Moreover, we adopt this criterium here because it was applied in A18 and we intend to compare 
the homogeneity scale obtained here (using the SDSS blue galaxies) with that one from A18 (using 
the ALFALFA catalogue).

The 1\%-criterium establishes the following procedure to identify the  homogeneity scale. 
First, the $\mathcal{D}_{2}(\theta)$ data points, calculated as described in previous subsection, are fitted by a model-independent polynomial\footnote{The order of this polynomial fit is chosen to minimise the root mean square error, taking into account the Akaike Information Criterium.}. 
The $\theta_{\text{H}}$ (or $r_{\text{H}}$ for a 3D distribution) is then defined by the numerical value at which this fitted curve attains the 1\% value below the fractal correlation dimension expected for the ideal case of a homogeneous distribution. 
%
%The 1\%-criterium establish that $\theta_{\text{H}}$ for a 2D distribution on a sphere, or $r_{\text{H}}$ for a 3D distribution, corresponds to the numerical value at which the function $\mathcal{D}_{2}$ attains the 1\% value below the correlation dimension expected for the ideal case of a homogeneous distribution. 
For a 3D distribution this means that $r_{\text{H}}$ is such that\footnote{For a discussion about the possibility to use $r_{\text{H}}$, defined by $\mathcal{D}_{2}(r_{\text{H}}) = 2.97$, as a cosmic standard ruler, similar to those performed with the BAO scale, and its use for cosmological parameter analyses, see \cite{Ntelis2018, Nesseris2019, Ntelis2019}. 
%Alternatively, one can find the best set of  cosmological parameters that provides an expected  homogeneity scale \citep{Scrimgeour} close to the measured scale through a Bayesian analysis (or similar).
}: 
$\mathcal{D}_{2}(r_{\text{H}}) = 2.97$. 
For a 2D distribution of objects on a sphere, $\mathcal{S}^{2}$, we have (see A18 for details)

%\begin{equation}\label{eq6}
%\mathcal{D}_{2}(\theta_{\text{H}}) \,=\, 0.99 \,\, 
%[\dfrac{\theta_{\text{H}}~\sin\theta_{\text{H}}}{1 - \cos\theta_{\text{H}}} \, .
%\end{equation}
%
\begin{equation}\label{eq6}
\mathcal{D}_{2}(\theta_{\text{H}}) \,=\, 0.99 \,\, 
\bigg [\dfrac{\theta~\sin\theta}{1 - \cos\theta} \bigg ]_{\theta = \theta_{\text{H}}} \, ,
\end{equation}
where the term in brackets $[\ast]$ corresponds to the fractal correlation dimension expected for a homogeneous distribution.
One interesting feature of this criterium is that it does not depend on the observational features of the survey \citep{Scrimgeour}, which makes suitable for the comparison of outcomes from different surveys and cosmic tracers, like the HI extragalactic sources and the SDSS blue galaxies. 

%Finally, to numerically obtain the $\theta_{\text{H}}$ we use a model-independent polynomial fit to describe the data points calculated in the previous section. 
%The order of this polynomial fit is chosen to minimize the root mean square error, and taking into account the Akaike Information Criterium. 

%The angular homogeneity scale, $\theta_{\text{H}}$, is obtained adopting a 
%well-established by~\cite{Scrimgeour} (where they arrive to theoretical and observational 
%arguments to decide the best criterium), 
%The adopted criterium to determine the angular scale, $\theta_{H}$, where the distribution can 
%be considered homogeneous, is the 1\% criterium discussed and adopted in~\cite{Scrimgeour}, 

%%----------------------------------------------------------------------------------------------------------------------
\section{Results}\label{sec4}

According to the main objectives exposed in the section~\ref{sec1}, we present the results of our 
analyses. 
We begin showing the analysis to determine the angular scale of transition to homogeneity, 
$\theta_{\text{H}}$, in the selected sample of SDSS blue galaxies. 
This is followed by our estimates of the relative bias among SDSS blue galaxies and ALFALFA 
sources and, finally, by the robustness tests performed to verify the validity of ours results.

%%----------------------------------------------------------------------------------------------------------------------
\subsection{The angular scale of transition to homogeneity}

\begin{figure}
\centering
\includegraphics[width=\columnwidth]{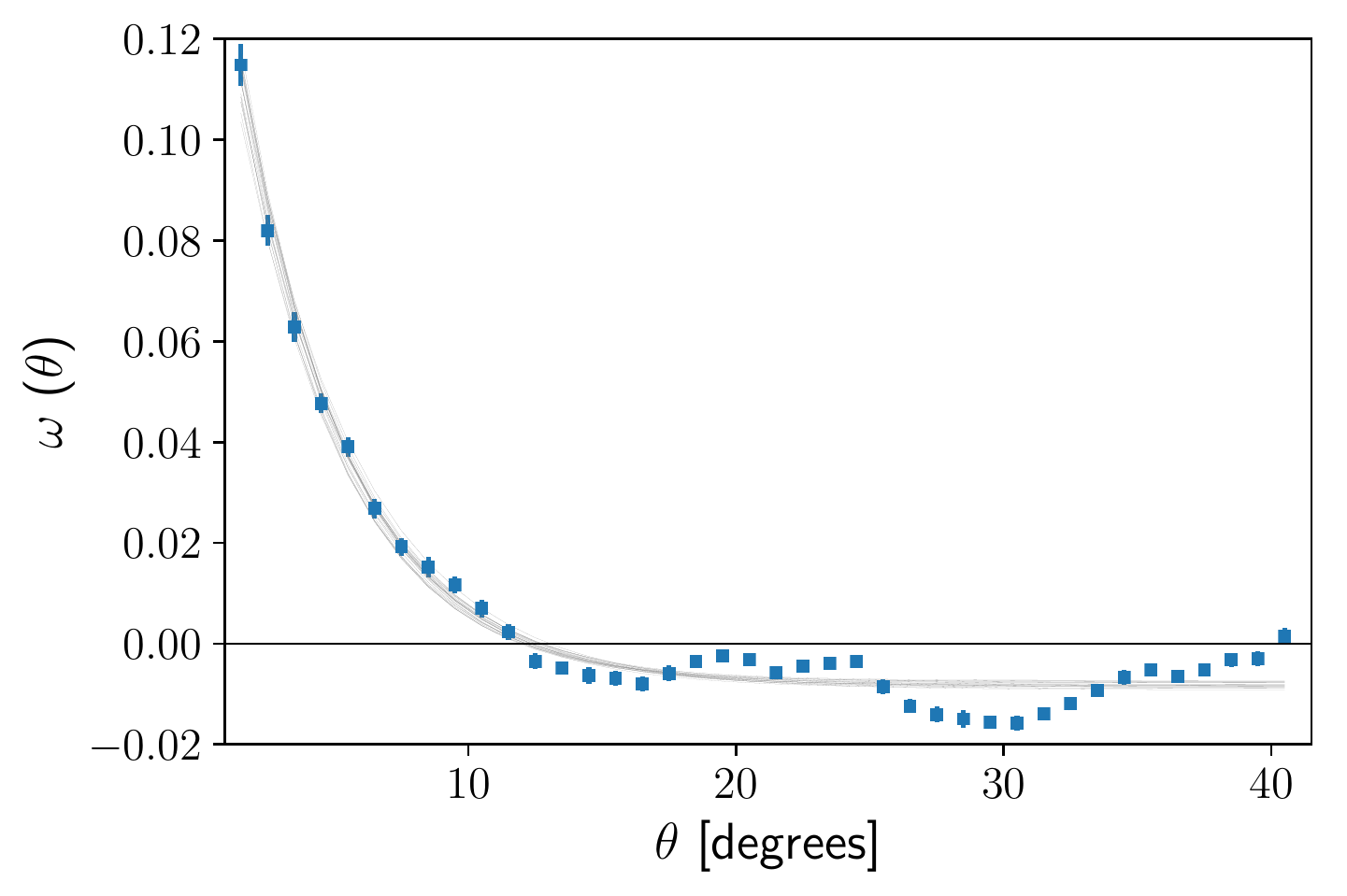}
%\vspace{-0.5cm}
\caption{Two-point angular correlation function for the selected sample of SDSS blue galaxies. 
The blue dots and error bars correspond to the average and standard deviation from the set of $\{\omega_j(\theta), j = 1, ..., 20\}$. 
The grey lines correspond to best-fit of $\omega_j(\theta)$ data points obtained using each of the 20 random catalogues.}
\label{fig4}
\end{figure}

We start the analyses producing a set of 20 simulated random catalogues, with the same 
observational features (sky area and number of objects) as the blue galaxies dataset, but uniformly 
distributed on the observed sky patch. 
In a previous work, A18, we used statistical tools to evaluate the performance of these random catalogues in homogeneity analyses, and confirmed its usability. 
With these catalogues we perform the calculation of $\omega_j(\theta)$, that is, the two-point 
angular correlation function for the $j$th random catalogue. 
In figure~\ref{fig4} we plot the best-fit curves of the data points from the set of $\{ \omega_j(\theta); j = 1,\cdots, 20 \}$ as continuous grey lines; the arithmetic mean of these data points, $\omega(\theta) = (1/20) \sum_{j=1}^{20} \omega_j(\theta)$, is shown, for illustrative purpose, as blue dots. 
Note the behaviour of the blue dots, oscillating below and above the best-fit curves.
 Given that these features appear below zero, this behaviour is probably reflecting the presence of different levels of under-densities (see also the discussion in A18).
%
%over- and under-dense regions in the SDSS blue galaxies dataset, as extensively discussed and tested in A18.

\begin{figure}
%\centering
\hspace{-0.15cm}
\includegraphics[width=\columnwidth]{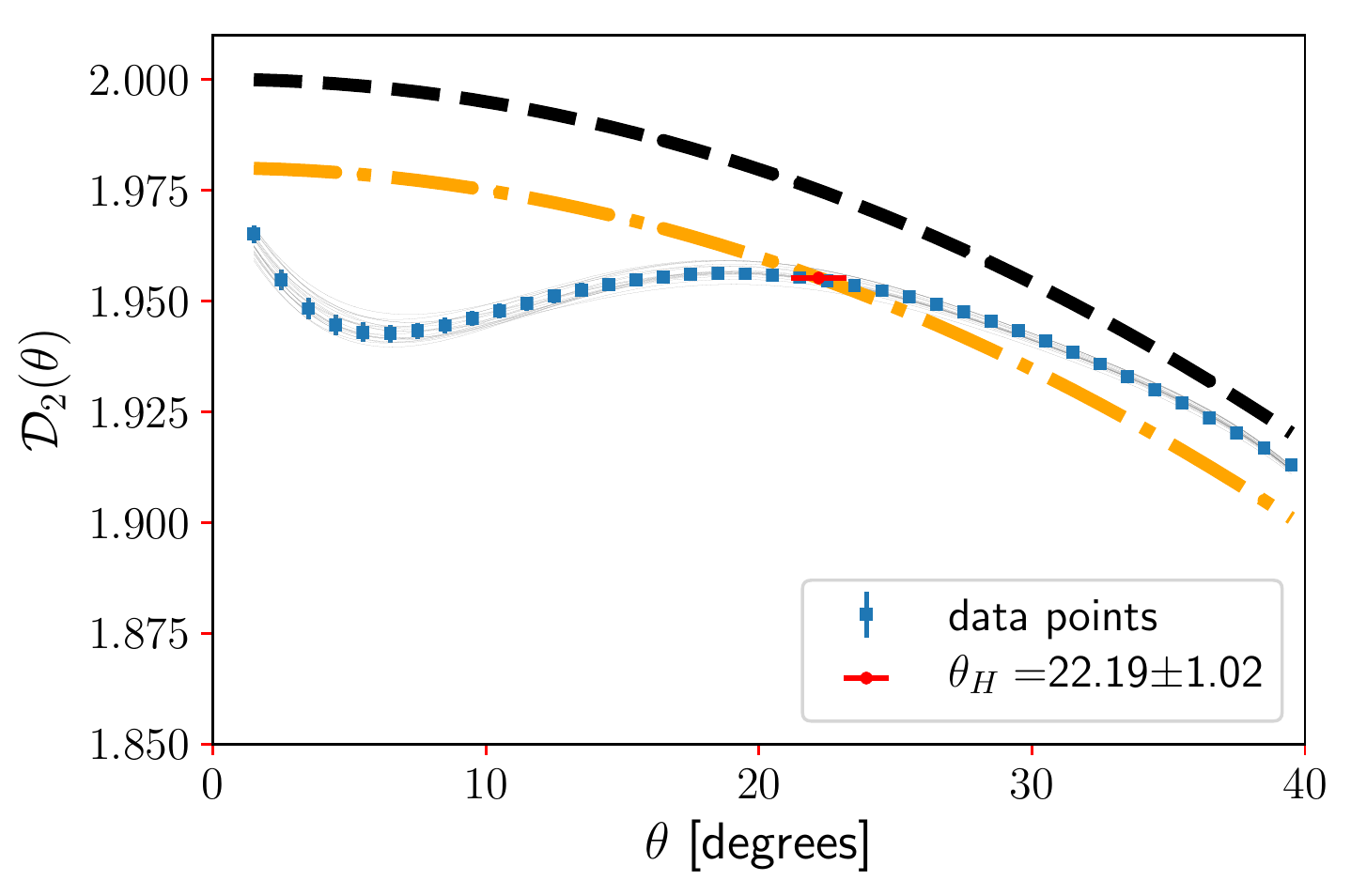}
%\vspace{-0.5cm}
\caption{Polynomial fits of order five (grey curves) of the $\mathcal{D}_{2}^j(\theta)$ data points measured using each of the 20 random catalogues through equation (\ref{eq2}). 
Their intersection with the orange dot-dashed line determines the angular scale of transition to homogeneity for each random catalogue, $\theta^j_{\rm H}$, whose average and standard deviation furnishes the measured $\theta_{\text{H}}$ value.
The average $\mathcal{D}_{2}(\theta)$ data points (blue dots) are presented for illustrative purpose. The black dashed line represents the threshold value for $\mathcal{D}_{2}(\theta)$ when the 
distribution is considered homogeneous (described by the term in brackets $[\ast]$ in equation \ref{eq6}). 
The orange dot-dashed line represents the 1\% below this threshold.} 
%
%$\mathcal{D}_{2}(\theta)$ data points (blue dots) obtained from equation~(\ref{eq2}). 
%The black dashed line represents the threshold value for $\mathcal{D}_{2}(\theta)$ when the 
%distribution is considered homogeneous (described by the term in brackets $[\ast]$ in equation \ref{eq6}). 
%The orange dot-dashed line represents the 1\% below this threshold. 
%The gray curves correspond to polynomial fits of order five for the $\mathcal{D}_{2}^j(\theta)$ 
%data points measured using each of the 20 random catalogues, whose intersection with the 
%orange dot-dashed line determines the angular scale of transition to homogeneity, 
%$\theta_{\text{H}}$.

\label{fig5}
\end{figure}

Using the best-fit curves from $\omega_j(\theta)$, we perform the next steps described in 
section~\ref{sec3}, estimating the  $\theta^j_H$ value for the $j$th random catalogue. 
The results of our analyses of the SDSS blue galaxies are shown in figure~\ref{fig5}, from where we 
obtain the final angular scale of homogeneity: $\theta_{\text{H}} = 22.19^{\circ} \pm 1.02^{\circ}$. 
The $\theta^j_{\rm H}$ values are obtained from the intersection points between the grey continuous curves (best-fit) and the dot-dashed line, applying the $1\%$-criterium described above. 
The dashed line represents the function in brackets $[\ast]$ in equation \ref{eq6}.
Each grey curve represents a polynomial fit of degree five for each set of $\mathcal{D}_{2}^{j}(\theta)$. 
Then the final result, $\theta_{\text{H}}$, is the arithmetic mean of the set of values obtained from each polynomial fit, $\langle \theta^j_{\rm H} \rangle$, and the error is its standard deviation, $\sigma(\theta^j_{\rm H})$. 
%modificado por Armando (7/maio):
This result is presented  in Table \ref{table1}, where we include, for comparison, our estimate for the theoretical prediction for $\theta_{\text{H}}$ according to the $\Lambda$CDM model. 
%, as well as the expected value from simulations of the dark matter density fluctuations, as done by \cite{Alonso14,Alonso15}. 
To estimate this value, we follow \citet[][Equations 3.4 and 3.5]{Edilson} to obtain the two point angular correlation function, using the non-linear power spectrum calculated by CAMB \citep{camb}, using the flat $\Lambda$CDM parameters $(\Omega_M, \Omega_\Lambda, \Omega_b, h, \sigma_8, n_s) = (0.3, 0.7, 0.049, 0.67, 0.8, 0.96)$, and corrected by redshift space distortion 
\citep[see][equation 21]{Alonso14}. 
Then, the predicted scale $\theta_{\text{H}}$ is calculated by fitting the $\mathcal{D}_{2}(\theta)$ obtained using equations \ref{eq2} and \ref{eq3}. 
Notice that to perform this procedure we calculate in advance the bias of our sample of 
blue galaxies following the prescription detailed in~\citet{Cresswell}, 
obtaining $b_{\footnotesize \rm blue} = 0.88$. As we shall show in the next subsection, the expected value for the angular scale of  homogeneity, $\theta_{\text{H}}$, is in excellent agreement with our estimate of this scale 
analysing both the SDSS blue galaxies and the extragalactic HI sources (in the later case 
the concordance is achieved after the corresponding bias correction, see Table \ref{table1}).

%%------------------------------------------------------  TABLE  ------------------------------------------------------
\begin{table*}
\centering
\begin{tabular}{| c | c | c | c | c | c |}
\hline
& ~~~~SDSS blue gals.~~~~  & ~~~~ALFALFA HI~~~~ & ~~~~$\!b^2_{\,\mbox{\footnotesize blue\,/\,\sc{HI}}}$~~~~ 
&  ~~~~ALFALFA HI~~~~  
& ~~~~theoretical~~~~ \\ % & expected from
& measured & measured & measured & bias corrected 
& prediction \\ %& simulations
\hline
~~~~$\theta_{\text{H}}$~~~~ & $22.19^{\circ}\pm1.02^{\circ}$ & $\!16.49^{\circ} \pm 0.29^{\circ}$ & $1.17 \pm 0.11$ & $\!20.62^{\circ} \pm 0.38^{\circ}$ & {$22.06^{\circ}$} \\
%& $\sim 22^{\circ}$
\hline
%\multicolumn{6}{p{2.5cm}}{\,} 
\end{tabular}
\caption{In this table we summarise the results concerning our analyses of the angular scale of 
transition to homogeneity as measured by the $\mathcal{D}_{2}(\theta)$ (or SCC) estimator.
We also display the theoretical prediction for $\theta_{\text{H}}$, 
obtained following the approach explained in the text and using the bias value 
$b_{\rm blue} = 0.88$, calculated for our blue galaxies sample according 
to~\citet{Cresswell}.}
%in the last column, 
%
%, bias corrected for the blue galaxies by using the HI bias factor relative to 
%matter~\citep[$b_{\rm HI} = 0.48 - 0.68$ depending on the richness of the sample;][]{Basilakos} 
%and the measured relative bias (fourth column). 
% 
%as well as its} value expected according to simulations, done for dark matter fluctuations 
%\citep[$b=1$;][]{Alonso14,Alonso15} and bias corrected for blue galaxies according to the 
%purity of the sample~\citep{Basilakos}.
\label{table1}
%, and also the value expected according to simulations (done for matter fluctuations, $b=1$; 
%see~\cite{Alonso14,Alonso15}) which is $32.46^{\circ}$, but according to~\cite{Basilakos}, 
%extragalactic HI line sources have an anti-bias $b$ with respect to matter fluctuations, a factor 
%that varies between 0.48 -- 0.68 depending on the richness of the sample. 
%For this, we obtain an interval of expected values for the ALFALFA catalogue, as stated in this 
%Table. 
\end{table*}
%0.48 \times 32.46, 0.68 \times 32.46
%O resultado do bias relativo foi de 1.43 +- 0.11
%O resultado para a distribui��o gaussianizada foi de 1.37 +- 0.26

%%-----------------------------------------------------------------------------------------------------------------------------
\subsection{The relative bias} 

It is possible to relate the angular scale of homogeneity from different biased cosmic tracers. 
This is done by correcting for bias the scaled counts-in-caps: 
$\mathcal{N}^{(2)}(<\theta) - 1 = b_{t2/t1}^2 \, [\,\mathcal{N}^{(1)}(<\theta) - 1\,]$, where 
$b_{t2/t1}^2$ is the relative bias of tracer 2 relative to tracer 1, and the upper index in 
$\mathcal{N}^{(i)}$ denotes the tracer $i$. 
Then, the angular scale of homogeneity of the tracer 2 can be obtained knowing 
$\mathcal{N}^{(1)}$ and $\mathcal{D}_{2}^{(1)}$ from the tracer 1, and the corresponding relative 
bias $b_{t2/t1}^2$ \cite[see, e.g.,][]{Scrimgeour,Laurent}. 

\begin{figure}
\centering
\includegraphics[width=\columnwidth]{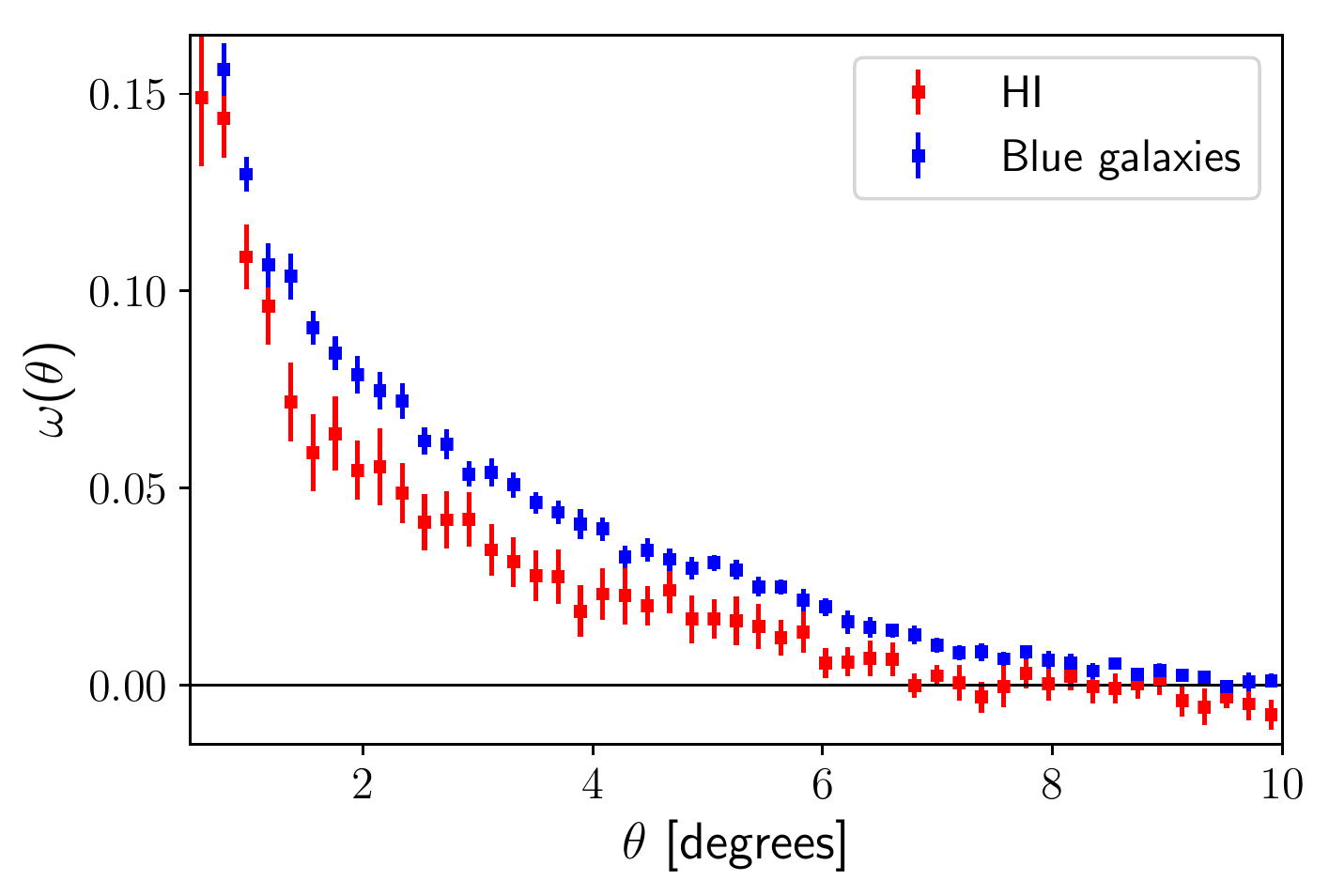}
%\vspace{-0.5cm}
\caption{Two-point angular correlation function calculated for the SDSS blue galaxies sample, and the ALFALFA sample analysed in A18. 
For each sample, the data points correspond to the average over the $\omega_j(\theta)$ calculated using each random catalogue. 
The graphic shows the $\theta$ range used to estimate the relative bias among the two samples.}
\label{fig6}
\end{figure}

The relative bias between the HI extragalactic sources from the ALFALFA catalogue and the SDSS 
blue galaxies, both samples containing sources in the local Universe $z < 0.06$, is obtained 
calculating the two-point angular correlation functions of the corresponding samples\footnote{Note that, it would be possible to estimate the bias by comparing $\theta_{\rm H}$ from each tracer. \cite{Ntelis2018} have shown this for $r_{\rm H}$, from the galaxy and the total matter distribution, by adapting the original linear bias model for $\omega(\theta)$ \citep[see also][]{Ntelis2019}.}: 
$b^2_{\,\mbox{\footnotesize blue\,/\,\sc{HI}}} \,\equiv\, 
\omega_{\,\mbox{\footnotesize blue}}(\theta) / 
\omega_{_{\mbox{\footnotesize \sc{HI}}}}(\theta)$~\citep{Scrimgeour,Papastergis}. 
In figure~\ref{fig6} we plot both correlation functions, where one observes that these samples 
have a relative bias close to one as expected~\citep{Papastergis}; in fact the numerical calculation gives: $b^2_{\,\mbox{\footnotesize blue\,/\,\sc{HI}}} = 1.17 \pm 0.11$.

%%----------------------------------------------------------------------------------------------------------------------------
\subsection{The robustness tests}\label{sec4.3}

One important part of our analyses is the realisation of robustness tests as a way to check for 
possible systematics affecting the outcomes \cite[see, e.g.,][]{Xavier18}. 

\begin{figure}
\centering
%$\mathcal{D}_{2}(\theta)$
\includegraphics[width=\columnwidth]{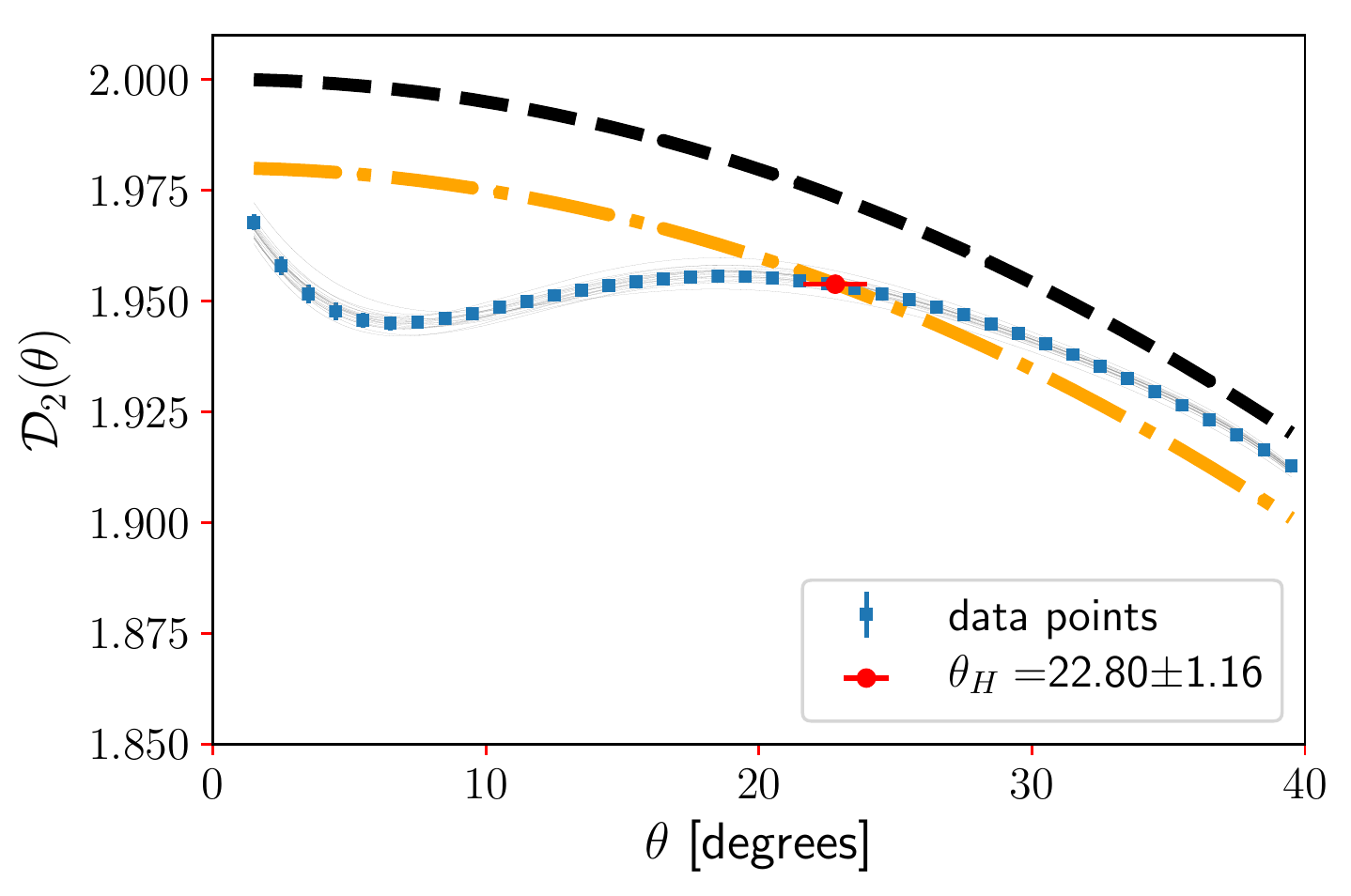}
%\vspace{-0.5cm}
\caption{Fractal correlation dimension analyses to find $\theta_{\text{H}}$ 
using twice the original sky area of the surveyed SDSS blue galaxies. 
This finding, $\theta_{\text{H}} = 22.80^{\circ} \pm 1.16^{\circ}$, agrees well with the results 
summarised in Table~\ref{table1} and shown in figure~\ref{fig5}.}
\label{fig7}
\end{figure}

The first item we study concerns the size of the surveyed area containing the data in analysis, 
where the question is if this area is large enough to accurately probe the angular scale of transition 
to homogeneity, $\theta_{\text{H}}$, we look for. 
To do this test, we shall consider a new sample of SDSS blue galaxies, namely, those displayed in 
an area twice the original one (shown in figure~\ref{fig1}). 
We select the region defined by the ranges 
$8^{\text{h}}40^{\text{m}} \leqslant \text{RA} \leqslant 16^{\text{h}}00^{\text{m}}$ and 
$0^{\circ} \leqslant \text{DEC} \leqslant 60^{\circ}$, obtaining a total 
of $57859$ blue galaxies. 
Then, we applied the same analyses described in section~\ref{sec3} over this new data sample, 
obtaining $\theta_{\text{H}} = 22.80^{\circ} \pm 1.16^{\circ}$, as  illustrated in figure~\ref{fig7}. 
This result confirms that the original area selected for the current analyses 
is suitable to investigate the transition to homogeneity of the SDSS blue galaxies in the local universe. 

\begin{figure*}
%\centering
\includegraphics[width=0.49\textwidth]{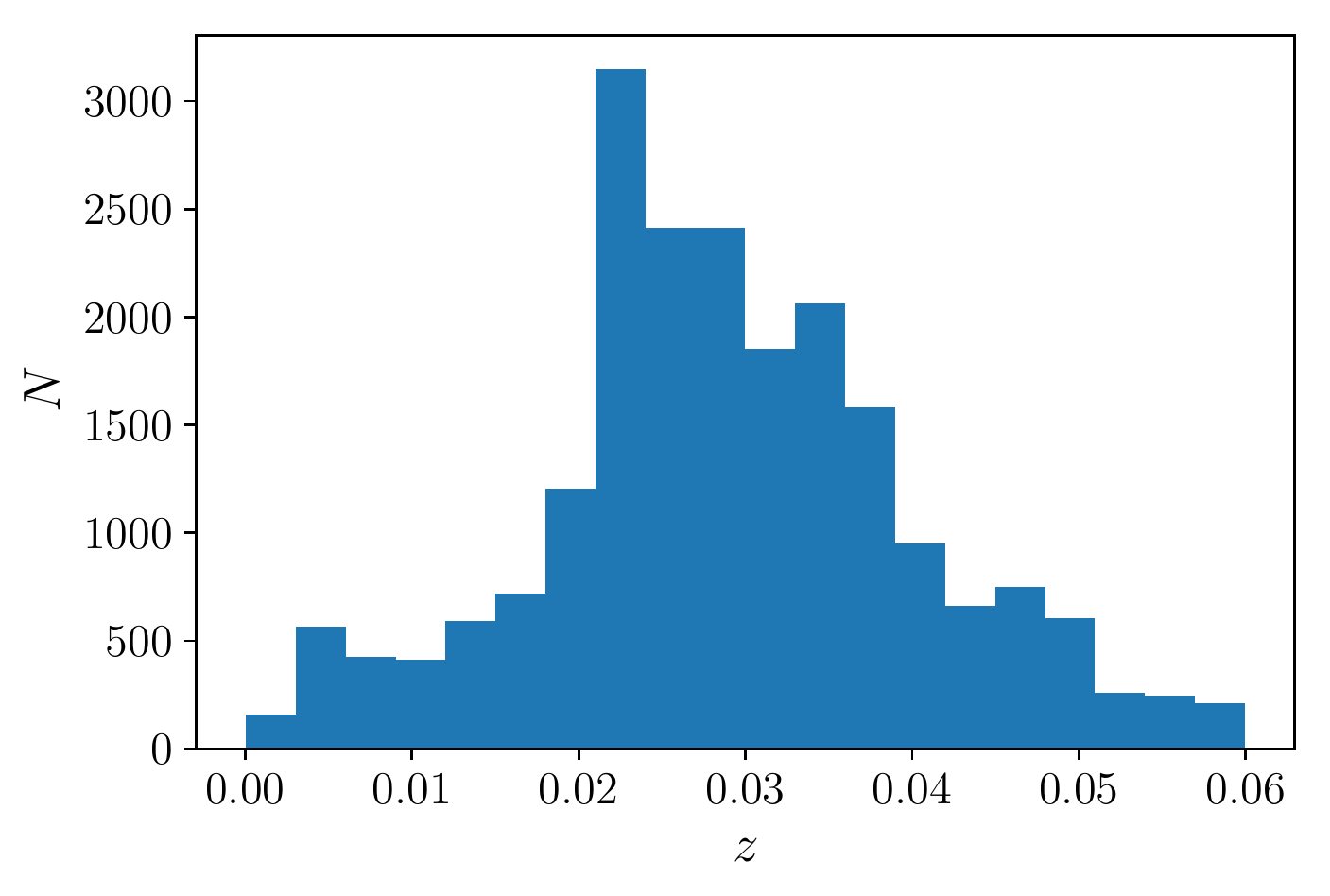}
\includegraphics[width=0.49\textwidth]{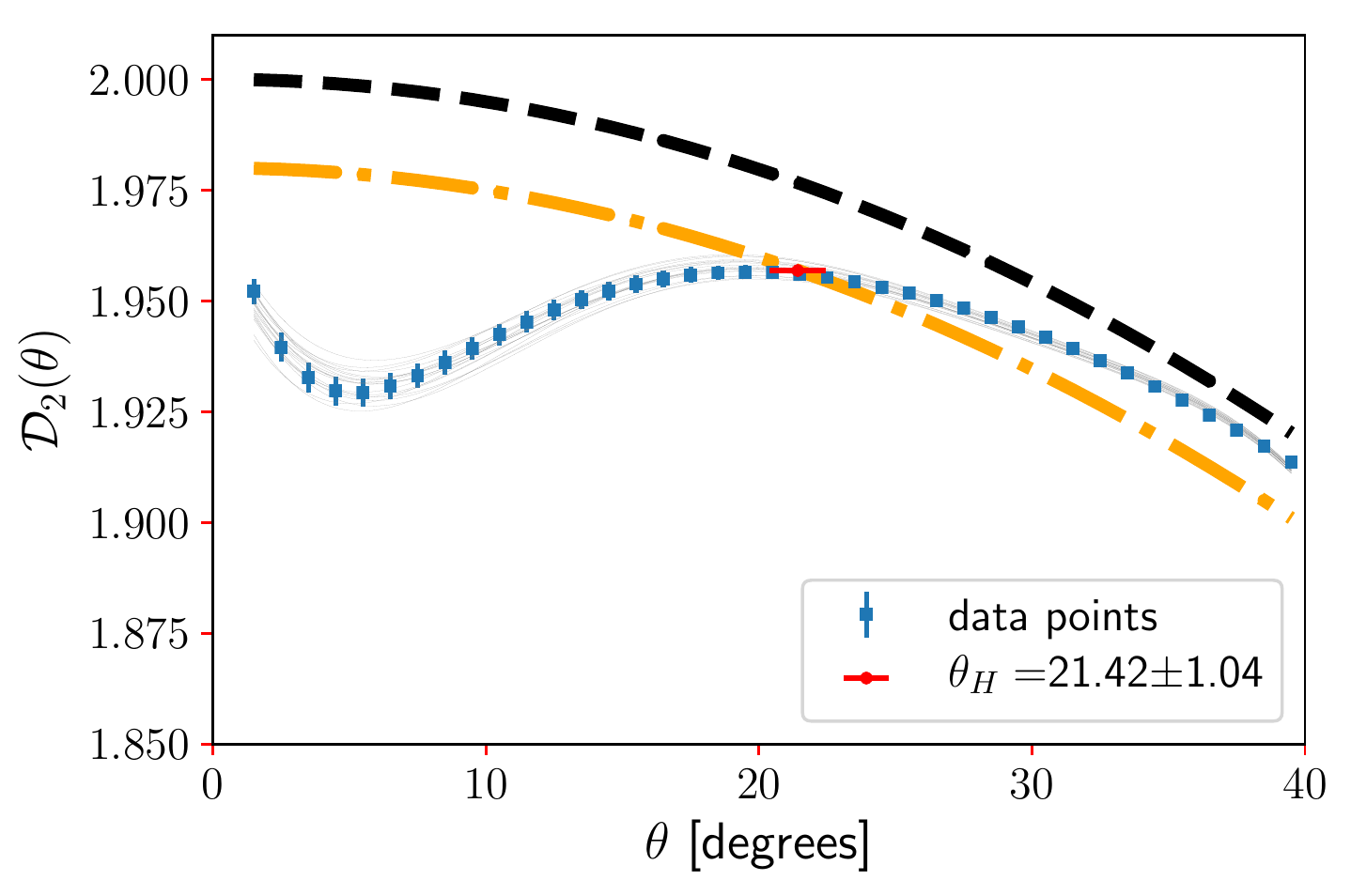}
%\vspace{-0.5cm}
\caption{
{\bf Left panel:} Redshift distribution of the reduced SDSS blue galaxies catalogue, with elements randomly removed in a way that the resulting sample has similar median and standard deviation parameters as those from the ALFALFA catalogue (see A18). 
{\bf Right panel:} Fractal correlation dimension analyses of the reduced sample whose distribution is shown in the left panel, which results in the angular scale of homogeneity $\theta_{\text{H}} = 21.42^{\circ} \pm 1.04^{\circ}$.}
\label{fig8}
\end{figure*}

A second effect with a potential influence on our results is the redshift distribution of our sample 
of SDSS blue galaxies. 
The redshift distribution of the sample in study, shown in figure~\ref{fig3} with median redshift 
equals to $0.037$, is different from that of the ALFALFA catalogue studied in A18 (with median 
redshift of $0.025$; see figure 2 there), with interesting results that we use for comparison here.
For this reason, we randomly remove blue galaxies in a way that the final sample has a 
median redshift and standard deviation close to those of the ALFALFA sample. 
The resulting sample contains $21226$ blue galaxies and can be observed in the left panel of 
figure~\ref{fig8}, with median redshift of $0.028$, from where we obtain 
$\theta_{\text{H}} = 21.42^{\circ} \pm 1.04^{\circ}$. 
This result is in excellent agreement with the results summarised in Table~\ref{table1}, making 
evident that our measurement of the angular scale of homogeneity in the local universe is robust. 

We also tested the robustness of our results with a different number of random catalogues, 
basically finding the same result. 
For the sake of comparison with the measurement from A18, the analyses presented here 
have been done using the same parameter choices as in that reference, namely, the number and 
size of $\theta$ bins, the number of random catalogues, and the 1\%-criterium.

%%----------------------------------------------------------------------------------------------------------------------------
\section{Summary and Conclusions}\label{sec5}

We investigate the clustering features of the blue galaxies sample observed by the SDSS in the 
local Universe ($0 < z < 0.06$; with median redshift $z_m = 0.037$).
The 2D projection of this sample was analysed using the scaled counts in spherical caps, 
$\mathcal{N}(<\theta)$, and the corresponding fractal correlation 
dimension, $\mathcal{D}_{2}(\theta)$, as described in section~\ref{sec4}~\citep{Scrimgeour,Laurent}. 
Such analyses allowed us to confirm that there is an angular scale of transition to homogeneity in 
the local Universe as traced by blue galaxies, namely, 
$\theta_{\text{H}} = 22.19^{\circ} \pm 1.02^{\circ}$.

Recently, in ref. A18, homogeneity analyses of the HI extragalactic sources of the ALFALFA catalogue resulted in the angular scale of transition to homogeneity $\theta_{\text{H}} = 16.49^{\circ} \pm 0.29^{\circ}$. 
Considering that both analyses, the current and the A18 ones, were done with data displayed in the same volume of the local Universe, the standard cosmological scenario foresees that angular scales of transition to homogeneity measured with different tracers should be bias related \cite[see, e.g.,][]{Scrimgeour,Ntelis}. 
This motivates us to search for this relation by performing a bias correction procedure. 
For this aim, we first look for the relative bias among the two tracers, the HI extragalactic sources 
and the SDSS blue galaxies. 
This analysis helped us to confirm that the HI extragalactic sources and the blue galaxies 
are cosmic tracers with similar, but not identical, clustering features as revealed by their relative 
bias: $b^2_{\,\mbox{\footnotesize blue\,/\,\sc{HI}}} = 1.17 \pm 0.11$ 
\cite[see, e.g.,][]{Papastergis}. 
Then, following~\cite{Scrimgeour}, we perform the bias correction in the functions $\mathcal{N}$ 
and $\mathcal{D}_{2}$ obtained from the ALFALFA sample to finally obtain the bias corrected measure 
of the angular scale of transition to homogeneity: 
$\theta_{\text{H}}^{\text{bias corr.}} = 20.62^{\circ} \pm 0.38^{\circ}$, in very good agreement with 
the value obtained in the analyses of the SDSS blue galaxies. 
Our comparison analysis is summarised in Table~\ref{table1}. 
We emphasise the importance of this agreement because it was obtained analysing two 
cosmological tracers, namely the HI sources from ALFALFA and the SDSS blue galaxies, observed 
in independent observational surveys with different instruments, systematics, and pipeline 
procedures~\citep{Haynes,York}.

We also performed robustness tests looking for potential sources of systematics that may affect 
the validity of our results. 
The first issue we investigate concerns the size of the surveyed area containing the SDSS data 
in analysis: the question is if this area is large enough to accurately probe the angular scale of 
transition to homogeneity, $\theta_{\text{H}}$. 
Our analyses explained in section~\ref{sec4.3} and displayed in figure~\ref{fig7} show that the 
original area containing the SDSS data, i.e., figure~\ref{fig1}, is large enough to probe the angular 
scale of transition to homogeneity.

Another important test concerns the potential influence of the redshift distribution of our sample 
of SDSS blue galaxies in the homogeneity analyses. 
In fact, the redshift distribution of the blue galaxies sample has median redshift equals to $0.037$ 
(see figure~\ref{fig2}) and does not seem to fit an approximate Gaussian distribution, as the 
redshift distribution of the ALFALFA catalogue studied in A18 does (with median redshift of $0.025$). 
To examine if this difference could be influencing our results we randomly remove blue galaxies 
in a way that the resulting sample has a median redshift and standard deviation close to those of 
the ALFALFA sample. 
After this process, the resulting sample contains $21226$ blue 
galaxies, with median redshift $0.028$ and distribution shown in the left panel of figure~\ref{fig8};
%galaxies and they can be observed in the left panel of figure~\ref{fig8}, with median redshift 0.028;
after analyses using this sample we observe in the right panel of figure~\ref{fig8} that 
$\theta_{\text{H}} = 21.42^{\circ} \pm 1.04^{\circ}$. 
This result is in excellent agreement with that one obtained with the original sample, making 
evident that our measurement of the angular scale of homogeneity in the local universe is robust.

Finally, it is worth emphasising that our analyses of the SDSS blue galaxies found an angular scale of transition to homogeneity in excellent agreement with the scale measured for the HI sample of the ALFALFA, 
%it is worth to mention that our analyses found an angular scale of transition to homogeneity for each cosmic tracer, however after a bias correction both scales are indeed in excellent agreement, 
corroborating the robustness of our study. 
Moreover, the angular scales obtained from the two analyses also agree well
%the angular scale of transition obtained in these analyses also agrees well 
with the value expected in the $\Lambda$CDM concordance model. 
%according to {\color{red} both, our theoretical prediction and the} simulations performed by~\cite{Alonso14,Alonso15} for dark matter (bias equal to 1), taking into account the individual bias of each cosmic tracer \citep{Basilakos,Papastergis}. 
Our comparative analyses of the angular scale of transition to homogeneity are summarised in Table~\ref{table1}.

%%----------------------------------------------------------------------------------------------------------------------------
\section*{Acknowledgements}

FA, CPN, and AB acknowledge fellowships from CAPES, FAPERJ, and CNPq, respectively. 
%Funda\c{c}\~ao de Amparo \`a Pesquisa no Amazonas
EdC acknowledges the PROPG-CAPES/FAPEAM program. 
We would like to thank Joel C. Carvalho, Rodrigo S. Gonçalves, Henrique Xavier, and Saulo Carneiro for useful comments and productive feedback of our analyses. 
We acknowledge the SDSS team for the use of the LSS data survey.

\end{document}